\documentclass[conference,a4paper]{IEEEtran}
\IEEEoverridecommandlockouts

\usepackage{cite}
\usepackage{amsmath,amssymb,amsfonts,dsfont}
\usepackage{algorithm,algorithmic}
\usepackage{graphicx}
\usepackage{textcomp}
\usepackage{xcolor}
\usepackage{bm}
\usepackage{booktabs}
\usepackage{multirow}
\usepackage{lipsum}
\usepackage{url}
\usepackage{amsthm}

\ifCLASSOPTIONcompsoc
\usepackage[caption=false, font=normalsize, labelfont=sf, textfont=sf]{subfig}
\else
\usepackage[caption=false, font=footnotesize]{subfig}

\newtheorem{remark}{Remark}

\newcommand{\thatis}{\textit{i.e.}}

\newcommand{\otw}{u}
 
\newcommand{\mbbR}{\mathbb{R}}

\newcommand{\bdw}{\boldsymbol{w}} 
\newcommand{\bdr}{\boldsymbol{r}}

\newcommand{\bdz}{\boldsymbol{z}}

\newcommand{\bdphi}{\boldsymbol{\phi}} 
\newcommand{\bdpsi}{\boldsymbol{\psi}} 

\newcommand{\bda}{\boldsymbol{\alpha}} 
\newcommand{\bdb}{\boldsymbol{\beta}} 
\newcommand{\bdlam}{\boldsymbol{\lambda}}

\newcommand{\mcX}{\mathcal{X}}

\newcommand{\mcY}{\mathcal{Y}}
\newcommand{\mcT}{\mathcal{T}}


\begin{document}

\title{Information Bottleneck Revisited: Posterior Probability Perspective with Optimal Transport
\thanks{This work was partially supported by National Key Research and Development Program of China (2018YFA0701603) and National Natural Science Foundation of China Grant (12271289 and 62231022).}
\thanks{The first two authors contributed equally to this work and the corresponding authors are Huihui Wu, Hao Wu and Wenyi Zhang.}
}

\author{\IEEEauthorblockN{Lingyi Chen\IEEEauthorrefmark{1},
Shitong Wu\IEEEauthorrefmark{1}\IEEEauthorrefmark{2},
Wenhao Ye\IEEEauthorrefmark{1}\IEEEauthorrefmark{2},
Huihui Wu\IEEEauthorrefmark{2},
Hao Wu\IEEEauthorrefmark{1},
Wenyi Zhang\IEEEauthorrefmark{3},
Bo Bai\IEEEauthorrefmark{2},
Yining Sun\IEEEauthorrefmark{2}
}
  
\IEEEauthorblockA{\IEEEauthorrefmark{1}Department of Mathematical Sciences, Tsinghua University, Beijing 100084, China}
\IEEEauthorblockA{\IEEEauthorrefmark{2}Theory Lab, 2012 Labs, Huawei Technologies Co., Ltd., Shatin, Hong Kong SAR, China }
\IEEEauthorblockA{\IEEEauthorrefmark{3}Department of Electronic Engineering and Information Science, \\ University of Science and Technology of China, Hefei, Anhui 230027, China, \\ Email: wu.huihui@huawei.com, hwu@tsinghua.edu.cn, wenyizha@ustc.edu.cn}
}

\maketitle

\begin{abstract}
Information bottleneck (IB) is a paradigm to extract information in one target random variable from another relevant random variable, which has aroused great interest due to its potential to explain deep neural networks in terms of information compression and prediction. Despite its great importance, finding the optimal bottleneck variable involves a difficult nonconvex optimization problem due to the nonconvexity of mutual information constraint. The Blahut-Arimoto algorithm and its variants provide an approach by considering its Lagrangian with fixed Lagrange multiplier. However, only the strictly concave IB curve can be fully obtained by the BA algorithm, which strongly limits its application in machine learning and related fields, as strict concavity cannot be guaranteed in those problems. To overcome the above difficulty, we derive an entropy regularized optimal transport (OT) model for IB problem from a posterior probability perspective. Correspondingly, we use the alternating optimization procedure and generalize the Sinkhorn algorithm to solve the above OT model. The effectiveness and efficiency of our approach are demonstrated via numerical experiments.
\end{abstract}


\section{Introduction} \label{sec_introduction}
The information bottleneck (IB) method, first proposed by Tishby et al. in 1999 \cite{tishby2000information_bottle,slonim2002ib_thesis}, was developed to extract the information of one target variable $Y$ from the observable variable $X$ in cases of unknown distortion measure \cite{book_element}.
Nowadays, it has a wide range of applications, such as information theory \cite{goldfeld2020ib_machine_learning,zaidi2020ib_view} and machine learning \cite{shamir2010learning_generalization,tishby2015deep_ib_principle,shwartz2017opening_box,saxe2019deep_learning}, because of its capability to evaluate the theoretical trade-off \cite{gilad2003ib_tradeoff} between complexity and predictive accuracy using information theoretic concepts.
Despite its great importance, finding the optimal bottleneck variable $T$ in the IB problem is not a simple task, due to the unconventional nonconvex constraint of the mutual information {\cite{boyd2004convex}}. 
To date, the most well-known numerical method for computing the IB problem is the Blahut-Arimoto (BA) algorithm \cite{tishby2000information_bottle,blahut1972computation}, which solves the Lagrangian of the IB problem with a fixed Lagrange multiplier (also called the IB Lagrangian) as an unconstrained objective function instead of solving the original IB problem as a constrained optimization problem. 
Under this framework, various methods based on similar idea of the BA algorithm has been proposed \cite{strouse2017dib,ni2022elastic_ib,alemi2016variation_ib,kolchinsky2019nonlinear_ib} to deal with the IB problem and its variants.
However, the solution of the IB problem is not always equivalent to the solution of its IB Lagrangian.
As it is shown in recent papers \cite{kolchinsky2018ib_caveats,kolchinsky2019nonlinear_ib}, when the IB curve is not strictly concave, there is no one-to-one mapping between the points on the IB curve and optimal solutions of the IB Lagrangian.
For example, in some supervised classification scenarios {\cite{amjad2019classification,goldfeld2019estimating,achille2018emergence}} where the predicted random variable is a deterministic function of the input observation variable, the IB curve contains a segment with a constant slope.
In these cases, the BA-type algorithm usually fails to work, since it can only obtain a few limited points rather than the entire IB curve, imposing a strong limit on its application in machine learning and related fields.
In this paper, we propose a novel approach to solve the IB problem directly as a constrained optimization problem rather than considering its IB Lagrangian. 
Specifically, we introduce {auxiliary optimization variables} to alleviate the challenges in numerical computation imposed by the mutual information constraint $I(T;Y)$ via solving the IB problem in a higher dimensional space.
Then, given the important observation that $I(X; T)\!=\!H(X)\!-\!H(X \!\mid\! T)$, where $H(X \!\mid\! T)$ is the effective term in the mutual information objective expressed by the posterior probability $P_{X\mid T}$, we propose to solve the IB problem from a posterior probability perspective.
Finally, we notice that the posterior probability $P_{X|T}$ and the conditional entropy function in the IB problem constitute a pair analogous to the pair of the transportation plan and objective function in the entropy regularized optimal transportation (OT) problem, as discussed in \cite{ye2022optimal,wu2022communication}.
Hence, we named this model as the IB-OT model.
To solve the proposed IB-OT model with high efficiency, we generalize the form of a recently introduced algorithm, \thatis, the Alternating Sinkhorn (AS) algorithm \cite{wu2022communication}, named herein as the Generalized Alternating Sinkhorn (GAS) algorithm. 
It is worth mentioning that the Lagrangian multipliers of the IB-OT model are updated during the iteration, in sharp contrast to the fixed multipliers in the BA algorithm \cite{kolchinsky2018ib_caveats,kolchinsky2019nonlinear_ib}. 
Moreover, by introducing {auxiliary optimization variables}, the Lagrangian of the IB-OT model is convex with respect to each primal variable, leading to better numerical stability when solving the subproblem in each alternating direction.
Additionally, closed form solution of primal variables can also be obtained in each alternating direction by considering of the posterior probability.
Since the IB problem is solved as a constrained optimization problem, the relevance-compression function can be obtained directly with a given threshold $I$.
Numerical experiments show that for classical cases like jointly Bernoulli and jointly Gaussian models, our proposed model and the GAS algorithm coincide with the theoretic results.
More importantly, our approach can overcome the limitations of the BA algorithm discussed in \cite{kolchinsky2018ib_caveats}, \thatis, even when the IB curve contains constant-slope segments, our approach can produce accurate numerical results, instead of only outputting a limited number of points like the BA algorithm. 
%

%
%
%
%
%
%
%


\section{Information Bottleneck Problem} \label{sec_2}
The information bottleneck (IB) is a method to extract information of a  target variable $Y$ from a correlated observable variable $X$ without using distortion measures. 
The extracted information is quantified by a bottleneck variable $T$, forming a Markov chain $Y \!\leftrightarrow\! X \!\leftrightarrow\! T$.
The IB problem is to find a bottleneck variable $T$ minimizing the mutual information $I(X ; T)$, while keeping the relevant information $I(Y ; T)$ above a certain threshold. 
Specifically, given a joint distribution $P_{XY}(x, y)$, the IB problem is defined as \cite{tishby2000information_bottle,goldfeld2020ib_machine_learning}
\begin{equation}\label{ibp_def}
    R(I):=\min _{P_{T \mid X}:~ I(T; Y) \geq I} I(X ; T).
\end{equation}
Here, $R(I)$ is called the relevance-compression function \cite{slonim2002ib_thesis} to represent the minimal achievable compression-information, for which the relevant information $I(Y ; T)$ is above $I$. 
Consider the case where $X$, $T$ and $Y$ are discrete random variables with input alphabet $\mcX=\{x_1,\cdots,x_M\}$, reproduction alphabet $\mcT=\{t_1,\cdots,t_N\}$ and output alphabet $\mcY\!=\!\{y_1,\cdots,y_K\}$.
Denoting $\otw_{ij}\!=\!P_{T\mid X}(t_j\!\mid\! x_i)$, $z_{kj}\!=\!P_{YT}(y_k,t_j)$ and $r_{j}\!=\!P_{T}(t_{j})$, according to the following formulas
\begin{equation*}
\begin{aligned}
    & P_{T}(t_{j})=\sum_{i=1}^{M} P_{T\mid X}(t_{j}\mid x_{i})P_{X}(x_{i}), \\
    & P_{YT}(y_{k},t_{j})=\sum_{i=1}^{M}P_{T\mid X}(t_{j}\mid x_{i})P_{Y\mid X}(y_{k}\mid x_{i})P_{X}(x_{i}),
\end{aligned}
\end{equation*}
the discrete form of IB problem \eqref{ibp_def} can be written as
\begin{subequations} \label{OT_model}
\begin{align}
&\min_{\bm{\otw},\bm{z},\bdr} && \sum_{i=1}^{M} \sum_{j=1}^{N} (\otw_{i j}p_{i}) \left(\log \otw_{i j}-\log r_{j}\right) \label{OT_model_a} \\
&\text{ s.t.} &&\sum_{j=1}^{N} \otw_{i j}p_{i}=p_{i}, \quad \sum_{i=1}^{M} \otw_{i j} p_{i}=r_{j}, ~ \forall i,j, \label{OT_model_b} \\
& &&\sum_{j=1}^{N} r_{j}=1, \quad \sum_{i=1}^{M}\otw_{ij}p_i s_{ki}=z_{kj}, ~ \forall j,k, \label{OT_model_c}\\
& &&\sum_{j=1}^{N}\!\left[\sum_{k=1}^{K}\sum_{i=1}^{M}\otw_{ij}s_{ki}p_i\log z_{kj}\!-\!r_j\log r_j\right]\! \geq \hat{I}. \label{OT_model_d}
\end{align}
\end{subequations}
Here, $p_{i}=P_{X}(x_{i})$, $q_{k}=P_{Y}(y_{k})$, $s_{ki}=P_{Y\mid X}(y_k\mid x_i)$ and  $\hat{I}=I+\sum_{k=1}^{K}q_{k}\log q_{k}$ are predetermined parameters.
Different from the BA algorithm, the {auxiliary variables} $\bdz$ and $\bdr$ are introduced explicitly as constraints \eqref{OT_model_b}, \eqref{OT_model_c} in our model to obtain a linear structure in a higher dimension space.
In fact, from the perspective of optimization, it is a normal practice to solve the optimization problem in a higher dimensional space \cite{sun2006optimization}.

\begin{remark}\label{lemma_1}
It is natural to express $I(T; Y)\geq I$ as a constraint in the following form 
%
\begin{equation}\label{newOT_model_d}
    \sum_{j=1}^N\sum_{k=1}^K z_{kj}(\log z_{kj}-\log r_j)\geq \hat{I}.
\end{equation}
By substituting $z_{kj}$ in \eqref{newOT_model_d} with the linear representation in \eqref{OT_model_b} and \eqref{OT_model_c}, we can obtain \eqref{OT_model_d}. However, replacing the term \eqref{OT_model_d} with \eqref{newOT_model_d} leads to the non-convexity in its Lagrangian function with respect to the direction in optimizing $z$.
\end{remark}

It is easy to verify that the Lagrangian function of \eqref{OT_model} is convex to each variable. 
However, based on this Lagrangian function, there is no analytical solution when updating $\bdr$. 
As a consequence, $\bdr$ needs to be solved numerically in its alternating direction, which brings numerical instability and high computational cost to the algorithm. 
Therefore, further exploration is required to find a better model, keeping the convexity with respect to each variable as well as a closed form solution when solving the subproblems by first order condition.
%


\section{Information Bottleneck Problem formulation from the Posterior Probability Perspective} \label{sec_3}
To deal with the above problem, we turn back to the decomposition of mutual information $I(X; T)\!=\!H(X)\!-\!H(X\!\mid\! T)$, where $H(X)$ is predetermined and $H(X\!\mid\! T)$ is the effective term in the objective function.
In this way, optimizing the objective of mutual information is reduced into optimizing the conditional entropy, which can be described by the posterior probability $P_{X\mid T}$.
Thus, based on this observation, we introduce the posterior probability $w_{ij}\!=\!P_{X\!\mid\! T}(x_i\!\mid\! t_j)$ to replace the prior probability $\otw_{ij}$ and propose the following model
\begin{subequations} \label{IB_OT_model}
\begin{align}
&\min_{\bdw,\bdr,\bdz} && \sum_{i=1}^{M} \sum_{j=1}^{N} w_{i j}r_j\log w_{i j} \label{IB_OT_model_a} \\
&\text{ s.t.} &&\sum_{j=1}^{N} w_{i j}r_{j}=p_{i}, \quad \sum_{i=1}^{M} w_{i j}r_j=r_j, ~ \forall i,j, \label{IB_OT_model_b} \\
& &&\sum_{j=1}^{N} r_{j}=1, \quad \sum_{i=1}^{M}w_{ij}r_j s_{ki}=z_{kj}, ~ \forall j,k, \label{IB_OT_model_c} \\
& &&\sum_{j=1}^{N}\!\left[\sum_{k=1}^K \sum_{i=1}^{M}w_{ij}s_{ki}r_j\log z_{kj}\!\!-\!\! r_j\log r_j\right]\!\!\geq\!\hat{I}. \label{IB_OT_model_d}
\end{align}
\end{subequations}
We can see that the above model is closely related to the entropy regularized OT problem \cite{ye2022optimal}.
Specifically, the posterior probability can be viewed as the transport plan with marginal distribution constraint \eqref{IB_OT_model_b}.
The objective function \eqref{IB_OT_model_a} can be viewed as the entropy regularized cost function in OT problems.
Besides, the constraints \eqref{IB_OT_model_c}, \eqref{IB_OT_model_d} induced by the {auxiliary variables} $\bdr,\bdz$ can be viewed as additional constraints in the classical OT problems. 
Thus, we name this newly proposed model the IB-OT model.
Denoting $\bda\in\mbbR^{M},\bdb\in\mbbR^{N},\bdlam\in\mbbR^{KN},\eta\in\mbbR,\zeta\in\mbbR^{+}$ as the Lagrange multipliers, the Lagrangian of the IB-OT model is given by:\par
\vspace{-.22in}
\begin{small}
\begin{equation} \label{Lagrangian_used}
\begin{aligned}
&\mathcal{L}(\bdw,\bdr,\bdz; \bda, \bdb, \bm{\lambda}, {\eta},\zeta)=\sum_{i=1}^{M} \sum_{j=1}^{N} w_{i j}r_j\log w_{i j} \\
&+ \sum_{i=1}^{M} \alpha_{i} \bigg(\sum_{j=1}^{N} w_{i j}r_j-p_i\bigg)+\sum_{j=1}^{N}\beta_{j}\bigg(\sum_{i=1}^{M} w_{i j} r_j-r_{j}\bigg) \\
&+\sum_{j=1}^N\sum_{k=1}^K{\lambda_{kj}}\bigg(\sum_{i=1}^{M}w_{ij}r_j s_{ki}-z_{kj}\bigg) + {\eta}\bigg(\sum_{j=1}^{N} r_{j}-1\bigg)\\
&-\zeta \bigg( \sum_{j=1}^N\sum_{k=1}^K \Big(\sum_{i=1}^{M}w_{ij}s_{ki}r_j\Big)\log z_{kj}\!\!-\!\!\sum_{j=1}^N r_{j}\log r_{j} \!\!-\!\! \hat{I} \bigg).
\end{aligned}
\end{equation}%
\end{small}%

Obviously, by checking the second order condition, this Lagrangian is convex with respect to each primal variable.
Moreover, analytical solutions of the primal variables are available when solving the first order condition, thereby ensuring efficiency and accuracy when designing algorithms.
It is worth emphasizing that the Lagrangian of IB-OT model is different from the IB Lagrangian with fixed Lagrange multiplier of the BA algorithm. 
This ensures that we can flexibly calculate the solution of the IB problem with the given threshold $I$.
In the next section, we propose an alternating minimization algorithm based on these properties of the IB-OT model.


\section{The Generalized Alternating Sinkhorn Algorithm}  \label{sec_4}

In this section, we generalize the Alternating Sinkhorn (AS) algorithm proposed in \cite{wu2022communication} for the computation of the IB-OT model and name it the Generalized Alternating Sinkhorn (GAS) algorithm.
Here, we sketch the main ingredients of our algorithm, while the detailed derivations are ignored due to space limitation.  
%

%

%
\begin{itemize}
    \item[A.] Fix $\bdr, \bdz$ as constant parameters and then update $\bdw$ and the associated dual variables $\bda,\bdb,\zeta$ in an alternating manner.
    Using the the idea of the Sinkhorn algorithm \cite{cuturi2013sinkhorn}, we can update $\alpha_{i}$ and $\beta_{j}$ as follows\par
    \begin{small}
    \begin{equation*}
        \psi_{j} = 1 \Big/ \sum_{i=1}^{M} \Lambda_{ij} \phi_{i} , \quad \phi_{i} = p_i \Big/ \sum_{j=1}^{N} \Lambda_{ij} \psi_{j} r_{j}.
    \end{equation*}%
    \end{small}%
    where $\phi_{i}=\exp(-{\alpha_{i}}-{1}/{2})$, $\psi_{j}=\exp(-\beta_{j}-{1}/{2})$ and $\Lambda_{ij}=\exp \big(-\sum_{k=1}^{K} s_{k i}(\lambda_{k j} -\zeta \log z_{kj})\big)$.

    Further, we can apply Newton's method to find the root of the monotonic function $G(\zeta)$ on $\mbbR^{+}$, where \par
    \vspace{-.22in}
    \begin{small}
    \begin{multline*}
        G(\zeta) \triangleq-\Big(\sum_{j=1}^{N} r_{j} \log r_{j}+\hat{I}\Big)+\sum_{j,k=1}^{N,K}\bigg(\sum_{i=1}^{M}\phi_is_{ki} \\
        \times\exp\Big(-\sum_{k'=1}^{K} s_{k'i}(\lambda_{k'j}-\zeta \log z_{k'j})\Big)\bigg)\psi_jr_j\log z_{kj}=0.
    \end{multline*}%
    \end{small}%
    Then, $\bdw$ is updated by $w_{i j}=\phi_{i}\Lambda_{ij}\psi_{j}$.
    \item[B.] Fix $\bdw,\bdr$ as constant parameters and then update $\bdz$ and the associated dual variable $\bdlam$. 
    We update $\bdlam$ by a closed form solution, \thatis, $\lambda_{kj}=-\zeta$.
    Then, $\bdz$ is updated by 
    \begin{equation*}
    z_{kj} = \sum_{i=1}^{M} s_{ki}w_{ij}r_j.
    \end{equation*}
    \item[C.] Fix $\bdw,\bdz$ as constant parameters and then update $\bdr$ and the associated dual variable $\eta$. 
    Also, we updated $\eta$ by a closed form solution, \thatis, $\eta=\zeta \log\Big(\sum_{j=1}^N \tilde{r_j}\Big)$, where \par
    \vspace{-.1in}
    \begin{small}
    \begin{multline*}\label{rj}
    \tilde{r_{j}}=\exp \bigg(-\frac{1}{\zeta}\sum_{i=1}^M\Big(w_{ij}\log w_{ij}-\zeta \sum_{k=1}^K s_{ki}w_{ij}\log z_{kj}\\
    \!+ \alpha_i w_{ij}\!+ \beta_j w_{ij}+\!\sum_{k=1}^K s_{ki}w_{ij}\lambda_{kj}-\beta_{j}\!\Big)\!\!-\!\!1\bigg).
    \end{multline*}
    \end{small}
    Then, we update $\bdr$ by $r_j=\tilde{r_{j}}\exp(-\eta/\zeta)$.
\end{itemize}

For clarity, the proposed GAS algorithm is summarized in Algorithm \ref{alg:OT_ibp}.
The significant difference between the GAS algorithm for IB-OT model and the AS algorithm for rate distortion function \cite{wu2022communication} is the occurrence of the update of $\bdz$ in Part B, which stems from the introduction of the slackness variable $\bdz$ to overcome the mutual information constraint as described before. 
We need to note that due to the nonconvexity of the IB problem, neither the Lagrangian of our IB-OT model nor the IB Lagrangian of the BA algorithm is globally convex.
Therefore, neither of them can guarantee global convergence. 
On the other hand, according to the numerical experiments in the next section, our GAS algorithm performs good convergence. 
In the classical cases, it matches the analytical solution and has a high efficiency advantage over the BA algorithm. 
For those tasks where the BA algorithm has limitations, our GAS algorithm can also perform well.
\begin{algorithm}[ht]
	\caption{Generalized Alternating Sinkhorn (GAS)}
	\label{alg:OT_ibp}
	\begin{algorithmic}[1]
		\REQUIRE Distribution $p_{i},q_k$, Conditional probability $s_{ki}$ \\
        Maximum iteration number $max\_iter$.
		\ENSURE Minimal value $\sum_{i=1}^{M} \sum_{j=1}^{N} w_{i j}r_j\log w_{i j}$.
        \STATE \textbf{Initialization:} $\bdphi=\mathbf{1}_{M},\bdpsi=\mathbf{1}_{N},\zeta,\eta=1,\bdlam=-\mathbf{1}_{KN}$, \\
        $\bdr=\mathbf{1}_{N}/N,\bdw=\mathbf{1}_{MN}/M, \bdz=\mathbf{1}_{KN}/KN$
		\STATE Set $\Lambda_{ij} \gets \exp(-\sum_{k=1}^{K} s_{k i}(\lambda_{k j} -\zeta \log z_{kj}))$
		\FOR{$\ell = 1 : max\_iter$}
		\STATE $\psi_{j} \gets 1/\sum_{i=1}^{M}\Lambda_{ij}\phi_{i}$
		\STATE $\phi_{i} \gets p_i/\sum_{j=1}^{N}\Lambda_{ij}\psi_{j}r_{j}$
		\STATE Solve $G(\zeta) = 0$ for $\zeta\in\mbbR^{+}$ using Newton's method
		\STATE $\Lambda_{ij} \gets \exp(-\sum_{k=1}^{K} s_{k i}(\lambda_{k j} -\zeta \log z_{kj}))$
            \STATE $w_{ij}\gets\phi_{i}\Lambda_{ij}\psi_{j}$
		\STATE $\lambda_{kj}\gets-\zeta$
        \STATE $z_{kj}\gets \sum_{i=1}^{M}w_{ij}r_j s_{ki}$
		\STATE Update $\tilde{r_{j}}$ and set $r_j\gets\tilde{r_{j}}/(\sum_{j=1}^N \tilde{r_{j}})$ 
		\ENDFOR
		\STATE \textbf{end}
		\RETURN $\sum_{i=1}^{M}\sum_{j=1}^{N} \left(\phi_{i}\Lambda_{ij}\psi_{j}r_{j}\right)\log \left(\phi_{i}\Lambda_{ij}\psi_{j}\right)$
	\end{algorithmic}
\end{algorithm}

\begin{remark} \label{remark_2}
A stabilization technique similar to log-domain stabilization technique \cite{chizat2018scaling} can also be integrated into our algorithm to resolve possible numerical issues caused by large $\zeta$ during iterations. 
Here, we make the following modifications in Algorithm \ref{alg:OT_ibp}. Replace line 2 and line 7 by:
\begin{equation*}
 \Lambda_{i j}\gets\exp \bigg(\!-\!\!\sum_{k=1}^K s_{k i} \lambda_{k j}
    \!+\!\zeta\Big(\sum_{k=1}^K s_{k i} \log z_{k j}\!-\!Z^{max}_{j}\Big)\bigg),
\end{equation*}
where $Z^{max}_{j}=\max_{i}\big(\sum_{k=1}^K s_{k i} \log z_{k j}\big)$.
Accordingly, in line 6 and line 11, we take the following substitution:
\begin{equation*}
 \psi_j\rightarrow \psi_j\exp\big(-\zeta Z^{max}_{j}\big).
\end{equation*}
\end{remark}


\section{Numerical Results and Discussions} \label{sec_5}

This section evaluates the effectiveness and efficiency of the proposed IB-OT model using the GAS algorithm. 
We consider three experiments: the classical models, a specially constructed model, and a real world dataset.
All the experiments are conducted on a PC with 8G RAM, and one Intel(R) Core(TM) Gold i5-8265U CPU @1.60GHz.

\subsection{Experiments for Classical Distributions}
This subsection computes the relevance-compression function $R(I)$ of two classical models, \thatis, 
the jointly Bernoulli model and the jointly Gaussian model.
The explicit expressions of the IB problem in these cases can be found in \cite{zaidi2020ib_view}.

For the jointly Bernoulli model
\begin{equation*}
    R(I)=\log 2-H\big((u-e)/(1-2e)\big),
\end{equation*}
where $u$ satisfies $I\!=\!\log(2)\!-\! H(u)$ and $e\!\leq\! u\!\leq\! 1/2$. 
Here, $H(u)$ is the entropy function and $X \sim\text{Bernoulli}(1/2)$, $Y\sim\text{Bernoulli}(1/2)$ and $X\oplus Y\sim\text{Bernoulli}(e)$, where the notation $\oplus$ means the sum modulo $2$ and $0\leq e\leq 1/2$. 
Moreover, the computation can be conducted directly due to the discrete distribution. 
In the following experiments, we take the flip probability $e = 0.15$.

For the jointly Gaussian model
\begin{equation*}
    R(I)=-\frac{1}{2}\log\Big(\big((1+\mathrm{SNR})\exp(-2I)-1\big)/\mathrm{SNR}\Big).
\end{equation*}
Here, $X$ is the standard Gaussian random variable with zero-mean and unit variance, and $X=\sqrt{\mathrm{SNR}}\ Y+S$, where $S$ is the standard Gaussian random variable with zero-mean and unit variance. 
We first truncate the variables into an interval $[-M,M]$ and discretize the interval by a set of uniform grid points $\{x_{i}\}_{i=1}^{N}$
\begin{equation*}
x_{i}=-M+(i-1)*\delta,~\delta=\frac{2M}{N},~i=1,\cdots,N.
\end{equation*}
Then, the discretized distribution is denoted as $p_{i} = \rho(x_i)\delta$, where $\rho(x)$ is the Gaussian density function.
In the following experiments, we take $\text{SNR} = 1, M = 10, \delta = 0.2$.

\vspace{-.14in}
\begin{figure}[H]
    \centerline{\includegraphics[width=0.5\textwidth]{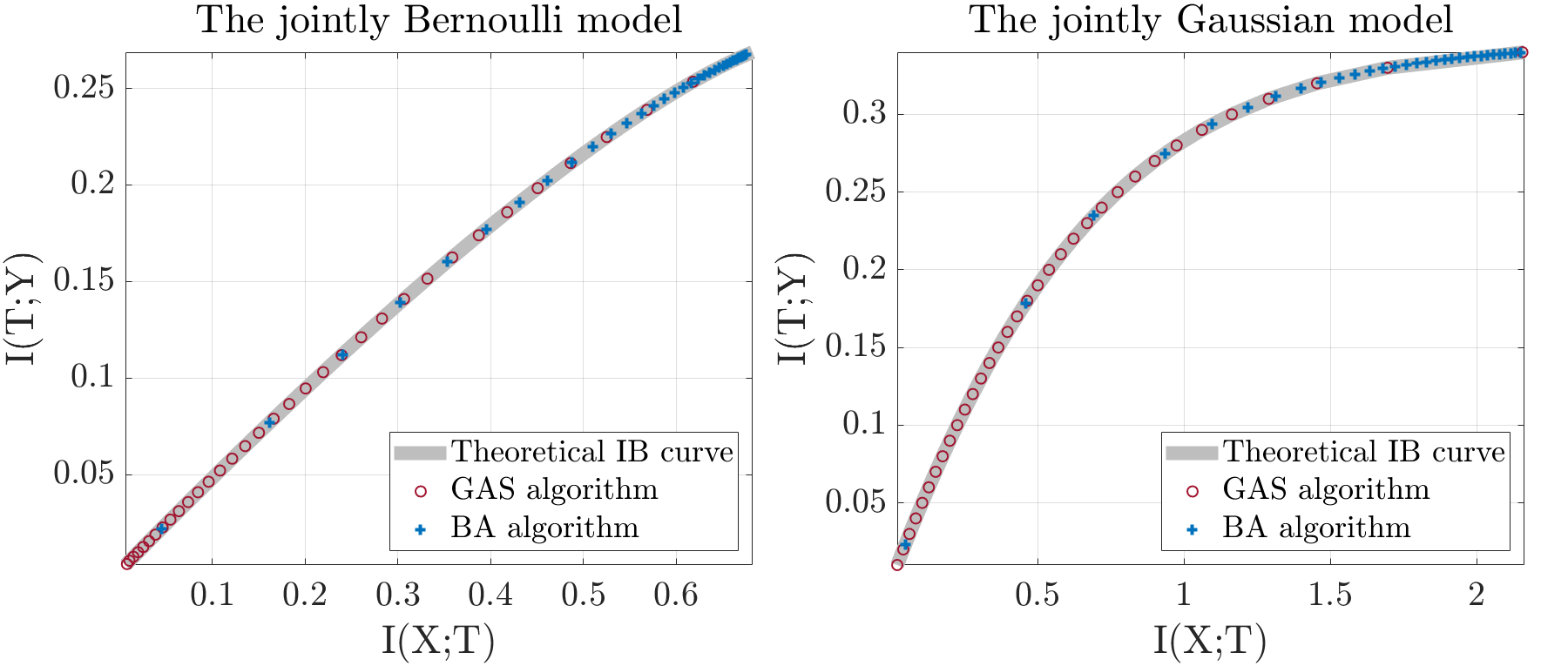}}
    \label{Fig: gas_bin_plot} 
    \caption{The comparison among analytical solution and GAS algorithm, BA algorithm for classical models.} 
\end{figure}
\vspace{-.11in}
In Fig. 1, we plot the IB curve given by the GAS algorithm and then compare the results with the theoretical IB curve as well as the BA algorithm. 
As shown in this figure, the GAS algorithm perfectly matches the relevance-compression functions $R(I)$ in both scenarios. 
Moreover, it is worth mentioning that in the Gaussian case, the relevance-compression curve given by the BA algorithm is still sparse for small $I(T; Y)$ values even though the same number of points are plotted. 
This phenomenon suggests the comparative advantage of the GAS algorithm in directly computing the relevance-compression functions.
\vspace{-.12in}
\begin{table}[ht] \label{table_compare} 
	\renewcommand\arraystretch{1.4}
	\centering
	\caption{Comparison between the GAS algorithm and the BA algorithm. } 
	\setlength{\tabcolsep}{2.4 mm}{
		\begin{tabular}{c|c|c|c|c}
			\toprule
			\multirow{2}{*}{} & \multirow{2}{*}{($I$, Slope $\zeta_{I}$)} & \multicolumn{2}{c|}{Time (s)}& \multicolumn{1}{c}{Ratio} \\
			\cline{3-5}
		      &  & $t_{GAS}$ & $t_{BA}$ & Speed-up \\ 
			\hline 
                \multirow{3}{*}{Bernoulli}
			& $(0.0823,2.1906)$ & $0.0063$ & $0.011$ & $1.75$ \\
			& $(0.1308,2.3299)$ & $0.0033$ & $0.0047$ & $1.42$ \\
                & $(0.1927,2.6432)$ & $0.0024$ & $0.0023$ & $0.96$ \\
			\hline
			\multirow{3}{*}{Gaussian} 
			& $(0.1,2.5687)$ & $0.48$ & $39.59$ & $82.48$ \\
			& $(0.2,3.9357)$ & $0.83$ & $50.65$ & $61.02$ \\
                & $(0.3,11.2435)$ & $2.24$ & $124.11$ & $55.41$ \\
			\bottomrule
	\end{tabular}}
 
\vspace{+.03in}

\footnotesize{{Notes: a) Column 3-4 are the average computing time, and column 5 is the speed-up ratio between the GAS and BA algorithm. b) The BA algorithm cannot compute the rate directly with a given $I$, and hence we adaptively search the corresponding slope $\zeta_I$ to ensure accuracy. It generally takes about $30\sim 50$ trials to search for a suitable slope $\zeta_I$. c) We set the stop condition that the difference with the analytical solution is less than $10^{-6}$.}}
\end{table}

\vspace{-.01in}

To further illustrate the efficiency of the GAS algorithm, we compare its computational cost with the BA algorithm as the baseline. 
The average computing time of the two methods under different choices of threshold parameter $I$ is listed in Table I.
To obtain a stable result, we repeat each experiment for $100$ times. 
%
%
From the table, we can see our proposed GAS algorithm has a significant advantage in computing time for the jointly Gaussian model. 
As for the jointly Bernoulli model, since the case is simple, it is natural that the two algorithms are almost equally efficient.

\subsection{Convergence Behaviour and Algorithm Verification}
In this subsection, we verify the convergence of the GAS algorithm by considering the absolute error between the result obtained by the GAS algorithm and the analytical solution of the relevance-compression function.
Here, we compute the absolute error with different threshold parameter $I$ for the jointly Bernoulli model and the jointly Gaussian model.
\vspace{-.1in}
\begin{figure}[H]
    \centerline{\includegraphics[width=0.5\textwidth]{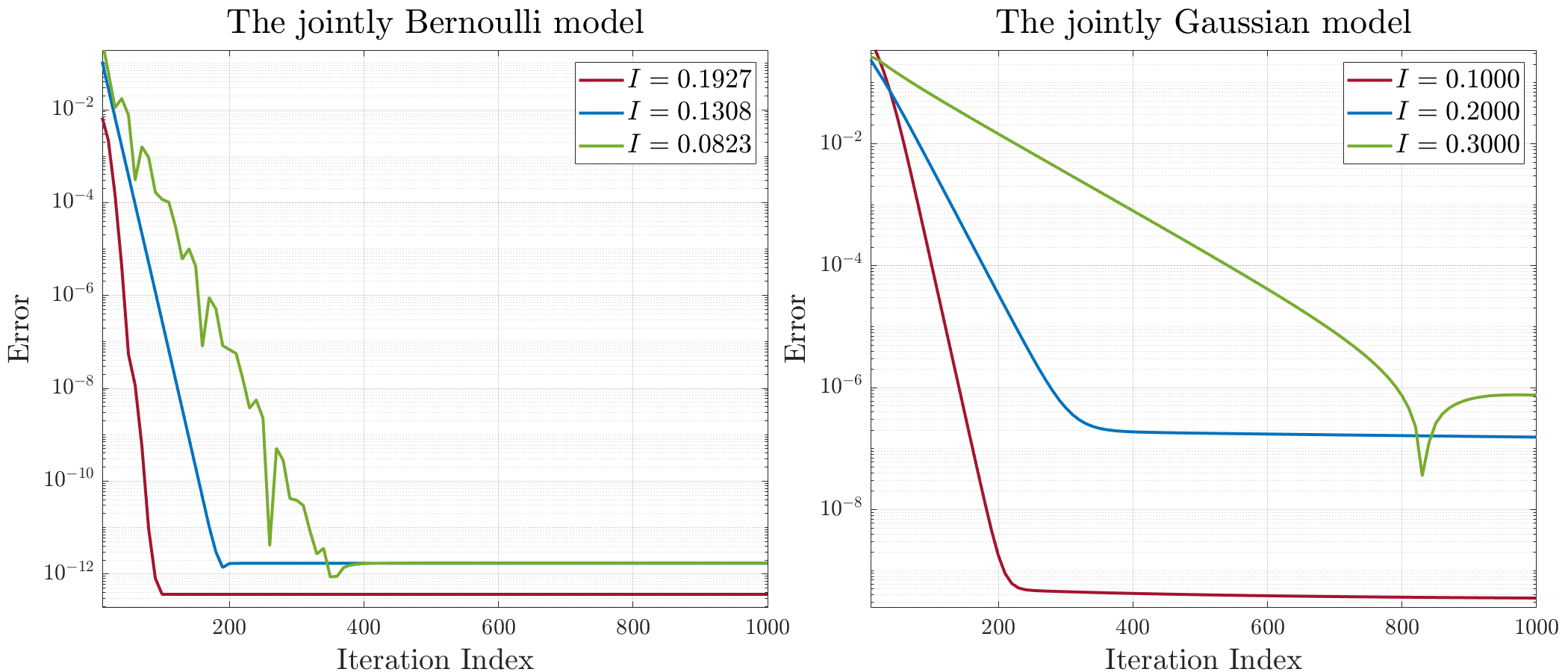}}
    \label{Fig: res_1}
    \caption{The convergent trajectories of the absolute error $r$ from the GAS algorithm for the classical models.
    }
\end{figure}
\vspace{-.1in}
In these two experiments, we set parameters the same as the parameters used for these two models in the above subsection and the maximum number of iterations is 1000.
As shown in Fig. 2, the GAS algorithm will successfully converge to $1e$-$6$ at last in all these cases.

\subsection{The Case of IB Curve with a Constant Finite Slope}

In this subsection, we consider a scenario where the IB curve contains a segment with a constant finite slope.
In this case, the theoretical IB curve is not strictly concave, so the BA algorithm cannot fully resolve the entire curve \cite{kolchinsky2018ib_caveats}.
Here, we consider an example of the joint distribution $P_{XY}(x,y)$ defined in \cite{gilad2003ib_tradeoff}, \thatis,\par
\vspace{-.1in}
\begin{small}
\begin{equation*}
P_{XY}(x, y)=\frac{1}{8}\left(\begin{array}{llll}
1 & 1 & 0 & 0 \\
1 & 1 & 0 & 0 \\
0 & 0 & 1 & 1 \\
0 & 0 & 1 & 1
\end{array}\right).
\end{equation*}
\end{small}

As shown in Fig. 3, the BA algorithm only outputs two points, no matter how initial values are selected.
On the other hand, our GAS algorithm could produce the entire IB curve and perfectly matches the analytic expression.
\vspace{-.1in}
\begin{figure}[H]                
    \centerline{\includegraphics[width=0.49\textwidth]{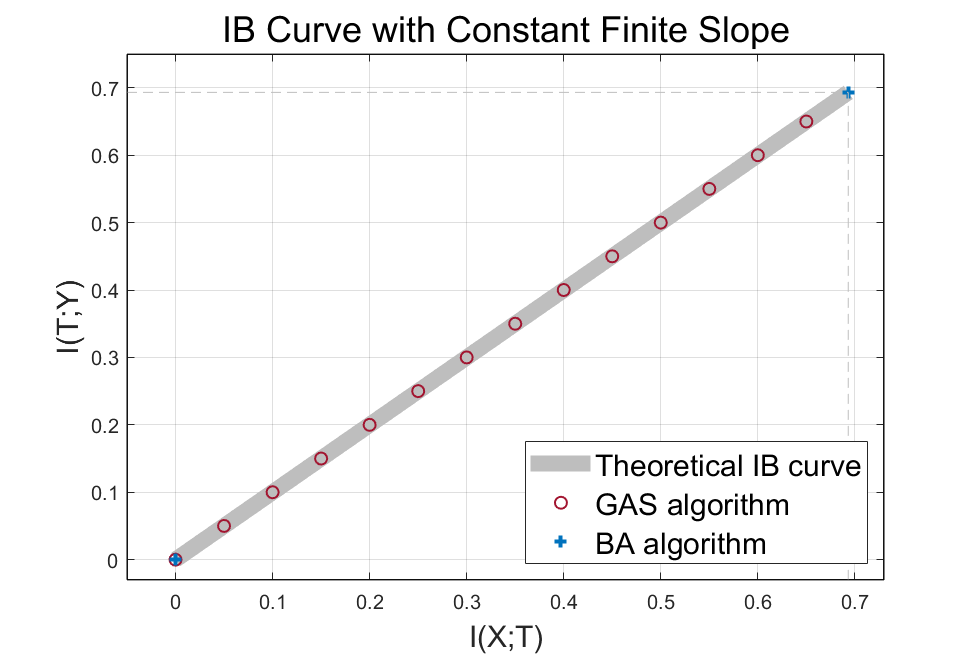}}
    \label{Fig: des_plot} 
    \caption{The comparison among theoretical IB curve, GAS algorithm and BA algorithm for a scenario with a constant finite slope.} 
\end{figure} 

\subsection{Iris Data Set -- A Real World Example}

In this subsection, we conduct the GAS algorithm using a real-world classification data set: the iris data set from the UCI learning repository \cite{Irisdata}. 
This data set consists of 50 samples from three classes and each sample owns four features. 
For the IB problem arising from this classification task, the observable variable $X$ represents the sample and the target variable $Y$ represents the class. 
The joint distribution $P_{XY}(x, y)$ is set as the empirical distribution produced from the data. 
\vspace{-.1in}
\begin{figure}[H]                
    \centerline{\includegraphics[width=0.49\textwidth]{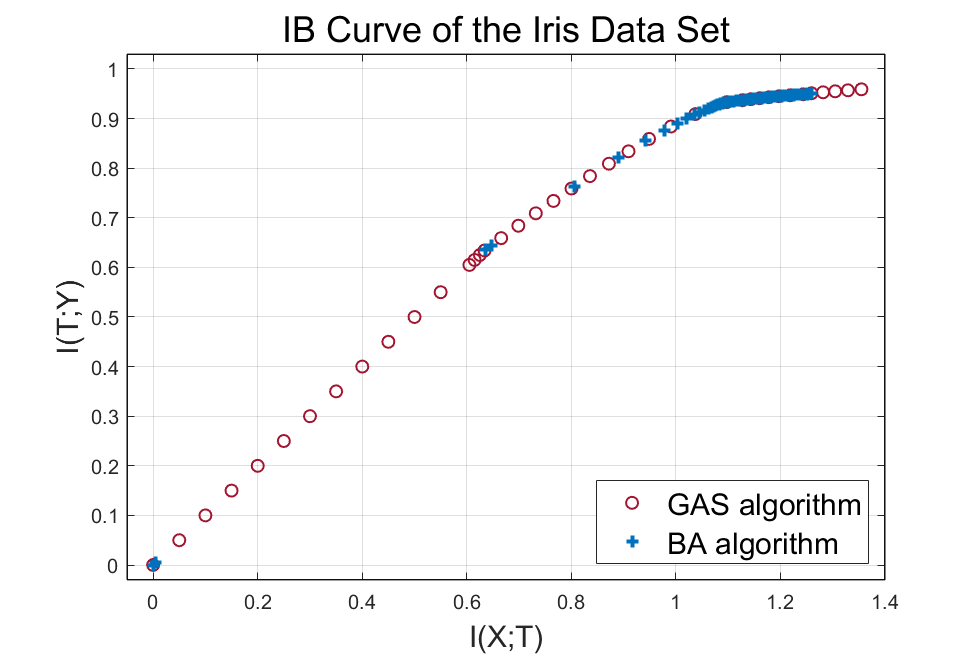}}
    \label{Fig: des_plot_2} 
    \caption{The comparison between GAS algorithm and BA algorithm for a scenario of real world data set for classification task.} 
\end{figure}
\vspace{-.1in}
As shown in Fig. 4, the IB curve corresponding to this task can be completely given by the GAS algorithm. 
For a given threshold $I$, the GAS algorithm can directly describe the trade-off between prediction and compression. 
%
%
{In contrast, the numerical results of many thresholds $I$ are missing by the BA algorithm.}
%

\section{Conclusion} \label{sec_6}
In this work, we present a novel framework for directly solving the IB problem. 
We reformulate the classic IB problem from a posterior probability perspective, through which we convert the original problem to a model with OT structure, \thatis, the IB-OT model. 
Furthermore, we propose the GAS algorithm by introducing {auxiliary optimization variables} to solve the optimization problem. 
Numerical experiments show that our algorithm is effective and efficient. 
Moreover, our results on the Iris Data Set and the deterministic task suggest the application potential of our approach to realistic machine learning tasks.

\bibliographystyle{bibliography/IEEEtran}
\bibliography{bibliography/IBP_REF}

\begin{thebibliography}{10}
\providecommand{\url}[1]{#1}
\csname url@samestyle\endcsname
\providecommand{\newblock}{\relax}
\providecommand{\bibinfo}[2]{#2}
\providecommand{\BIBentrySTDinterwordspacing}{\spaceskip=0pt\relax}
\providecommand{\BIBentryALTinterwordstretchfactor}{4}
\providecommand{\BIBentryALTinterwordspacing}{\spaceskip=\fontdimen2\font plus
\BIBentryALTinterwordstretchfactor\fontdimen3\font minus
  \fontdimen4\font\relax}
\providecommand{\BIBforeignlanguage}[2]{{%
\expandafter\ifx\csname l@#1\endcsname\relax
\typeout{** WARNING: IEEEtran.bst: No hyphenation pattern has been}%
\typeout{** loaded for the language `#1'. Using the pattern for}%
\typeout{** the default language instead.}%
\else
\language=\csname l@#1\endcsname
\fi
#2}}
\providecommand{\BIBdecl}{\relax}
\BIBdecl

\bibitem{tishby2000information_bottle}
N.~Tishby, F.~C. Pereira, and W.~Bialek, ``{The Information Bottleneck
  Method},'' in \emph{Proc. 37th Annual Allerton Conference on Communications,
  Control and Computing}, Monticello, Illinois, USA, Oct. 1999, pp. 368--377.

\bibitem{slonim2002ib_thesis}
N.~Slonim, ``{The Information Bottleneck: Theory and Applications},'' Ph.D.
  dissertation, Hebrew University of Jerusalem Jerusalem, Israel, 2002.

\bibitem{book_element}
T.~M. Cover and J.~A. Thomas, \emph{{Elements of Information Theory}}.\hskip
  1em plus 0.5em minus 0.4em\relax Wiley-Interscience, 2006.

\bibitem{goldfeld2020ib_machine_learning}
Z.~Goldfeld and Y.~Polyanskiy, ``{The Information Bottleneck Problem and Its
  Applications in Machine Learning},'' \emph{IEEE Journal on Selected Areas in
  Information Theory}, vol.~1, no.~1, pp. 19--38, Apr. 2020.

\bibitem{zaidi2020ib_view}
A.~Zaidi, I.~Estella-Aguerri, and S.~Shamai, ``{On the Information Bottleneck
  Problems: Models, Connections, Applications and Information Theoretic
  Views},'' \emph{Entropy}, vol.~22, no.~2, p. 151, Jan. 2020.

\bibitem{shamir2010learning_generalization}
O.~Shamir, S.~Sabato, and N.~Tishby, ``{Learning and Generalization with the
  Information Bottleneck},'' \emph{Theoretical Computer Science}, vol. 411, no.
  29-30, pp. 2696--2711, Jun. 2010.

\bibitem{tishby2015deep_ib_principle}
N.~Tishby and N.~Zaslavsky, ``{Deep Learning and the Information Bottleneck
  Principle},'' in \emph{Proc. 2015 IEEE Information Theory Workshop (ITW)},
  Jerusalem, Israel, Apr. 2015, pp. 1--5.

\bibitem{shwartz2017opening_box}
R.~Shwartz-Ziv and N.~Tishby, ``{Opening the Black Box of Deep Neural Networks
  via Information},'' \emph{arXiv preprint arXiv:1703.00810}, 2017.

\bibitem{saxe2019deep_learning}
A.~M. Saxe, Y.~Bansal, J.~Dapello, M.~Advani, A.~Kolchinsky, B.~D. Tracey, and
  D.~D. Cox, ``{On the Information Bottleneck Theory of Deep Learning},''
  \emph{Journal of Statistical Mechanics: Theory and Experiment}, vol. 2019,
  no.~12, p. 124020, Dec. 2019.

\bibitem{gilad2003ib_tradeoff}
R.~Gilad-Bachrach, A.~Navot, and N.~Tishby, ``{An Information Theoretic
  Tradeoff between Complexity and Accuracy},'' in \emph{Proc. 16th Annual
  Conference on Computational Learning Theory and 7th Kernel Workshop
  (COLT/Kernel 2003)}, Washington D.C., USA, Aug. 2003, pp. 595--609.

\bibitem{boyd2004convex}
S.~Boyd and L.~Vandenberghe, \emph{{Convex Optimization}}.\hskip 1em plus 0.5em
  minus 0.4em\relax Cambridge, UK: Cambridge University Press, 2004.

\bibitem{blahut1972computation}
R.~E. Blahut, ``{Computation of Channel Capacity and Rate-Distortion
  Functions},'' \emph{IEEE Transactions on Information Theory}, vol.~18, no.~4,
  pp. 460--473, Jan. 1972.

\bibitem{strouse2017dib}
D.~Strouse and D.~J. Schwab, ``{The Deterministic Information Bottleneck},''
  \emph{Neural Computation}, vol.~29, no.~6, pp. 1611--1630, Jun. 2017.

\bibitem{ni2022elastic_ib}
Y.~Ni, Y.~Lan, A.~Liu, and Z.~Ma, ``{Elastic Information Bottleneck},''
  \emph{Mathematics}, vol.~10, no.~18, p. 3352, Sep. 2022.

\bibitem{alemi2016variation_ib}
A.~A. Alemi, I.~Fischer, J.~V. Dillon, and K.~Murphy, ``{Deep Variational
  Information Bottleneck},'' in \emph{Proc. 5th International Conference on
  Learning Representations (ICLR)}, Toulon, France, Apr. 2017, pp. 1--5.

\bibitem{kolchinsky2019nonlinear_ib}
A.~Kolchinsky, B.~D. Tracey, and D.~H. Wolpert, ``{Nonlinear Information
  Bottleneck},'' \emph{Entropy}, vol.~21, no.~12, p. 1181, Nov. 2019.

\bibitem{kolchinsky2018ib_caveats}
A.~Kolchinsky, B.~D. Tracey, and S.~Van~Kuyk, ``{Caveats for Information
  Bottleneck in Deterministic Scenarios},'' in \emph{Proc. 7th International
  Conference on Learning Representations (ICLR)}, New Orleans, Louisiana, USA,
  May 2019, pp. 1--23.

\bibitem{amjad2019classification}
R.~A. Amjad and B.~C. Geiger, ``{Learning Representations for Neural
  Network-based Classification Using the Information Bottleneck Principle},''
  \emph{IEEE Transactions on Pattern Analysis and Machine Intelligence},
  vol.~42, no.~9, pp. 2225--2239, Apr. 2019.

\bibitem{goldfeld2019estimating}
Z.~Goldfeld, E.~Van Den~Berg, K.~Greenewald, I.~Melnyk, N.~Nguyen,
  B.~Kingsbury, and Y.~Polyanskiy, ``{Estimating Information Flow in Deep
  Neural Networks},'' in \emph{\emph{Proc.} 36th International Conference on
  Machine Learning (ICML)}, California, USA, Jun. 2019, pp. 4153--4162.

\bibitem{achille2018emergence}
A.~Achille and S.~Soatto, ``{Emergence of Invariance and Disentanglement in
  Deep Representations},'' \emph{The Journal of Machine Learning Research},
  vol.~19, no.~1, pp. 1947--1980, Sep. 2018.

\bibitem{ye2022optimal}
W.~Ye, H.~Wu, S.~Wu, Y.~Wang, W.~Zhang, H.~Wu, and B.~Bai, ``{An Optimal
  Transport Approach to the Computation of the LM Rate},'' in \emph{Proc. 2022
  IEEE Global Communications Conference (GLOBECOM)}, Rio de Janeiro, Brazil,
  Dec. 2022, pp. 239--244.

\bibitem{wu2022communication}
S.~Wu, W.~Ye, H.~Wu, H.~Wu, W.~Zhang, and B.~Bai, ``{A Communication Optimal
  Transport Approach to the Computation of Rate Distortion Functions},'' in
  \emph{\emph{Proc.} 2023 IEEE Information Theory Workshop (ITW)}, Saint-Malo,
  France, Apr. 2023.

\bibitem{sun2006optimization}
W.~Sun and Y.-X. Yuan, \emph{{Optimization Theory and Methods: Nonlinear
  Programming}}.\hskip 1em plus 0.5em minus 0.4em\relax Springer Science \&
  Business Media, 2006, vol.~1.

\bibitem{cuturi2013sinkhorn}
M.~Cuturi, ``{Sinkhorn Distances: Lightspeed Computation of Optimal
  Transport},'' in \emph{Proc. Advances in Neural Information Processing
  Systems (NIPS 2013)}, vol.~26, Lake Tahoe, Nevada, USA, Dec. 2013, pp.
  2292--2300.

\bibitem{chizat2018scaling}
L.~Chizat, G.~Peyr{\'e}, B.~Schmitzer, and F.-X. Vialard, ``{Scaling Algorithms
  for Unbalanced Optimal Transport Problems},'' \emph{Mathematics of
  Computation}, vol.~87, no. 314, pp. 2563--2609, Nov. 2018.

\bibitem{Irisdata}
C.~L. Blake and C.~J. Merz, ``{UCI Repository of Machine Learning Databases},''
  \emph{http://www.ics.uci.edu/mlearn/MLRepository.html}, 1998.

\end{thebibliography}

\end{document}